\documentstyle[aps,prl,multicol,epsfig]{revtex}
\begin{document}
\title{ Fractional charges and quantum phase transitions in imbalanced
       bilayer quantum Hall systems}
\author{  Jinwu Ye  }
\address{ Department of Physics, The Pennsylvania State University, University Park, PA, 16802 }
\date{\today}
\maketitle
\begin{abstract}
     We extend the Composite Boson theory to study slightly im-balanced
     bi-layer Quantum Hall systems. In the global $ U(1) $ symmetry breaking
     excitonic superfluid side, as the imbalance increases,
     the system supports continuously changing fractional charges.
     In the translational symmetry breaking pseudo-spin density wave (PSDW) side, there are two quantum
     phase transitions from the commensurate PSDW to an
     in-commensurate PSDW and then to the excitonic superfluid state.
     We compare our theory with experimental data and also the previous microscopic
     calculations.
%    A  scenario on how different
%    phases evolve as the distance between the two layers is increased is proposed.
\end{abstract}
\begin{multicols}{2}
   Spin-polarized Bilayer Quantum Hall systems at total filling factor
  $ \nu_{T} =1 $ have been under enormous experimental and theoretical investigations over the last decade
  \cite{rev}.  When the interlayer separation $ d $ is sufficiently large, the bilayer
  system decouples into two separate compressible $ \nu=1/2 $ layers
  \cite{supp}. However, when $ d $ is smaller than a critical distance $ d_{c1} $, even in the absence
  of interlayer tunneling, the system undergoes a quantum phase transition into
  a novel spontaneous interlayer coherent incompressible phase which
  is an excitonic superfluid state in the pseudospin channel.
  \cite{gold,hall,counterflow}.

    Although the ESF phase and FL phase at the two extreme distances are well established,
    the picture of how the ESF phase evolves into the two weakly-coupled FL states as distance changes
    is still not well established. The experiment \cite{drag}
    discovered that although there are very little dissipations
    in both the ESF and FL, there is a strong enhancement of drag and dissipations
    in an intermediate distance regime. These experimental observations
    suggest that there maybe an intermediate phase separating the two phases.
    If there is indeed such an intermediate phase, even it exists, the nature of the intermediate phase
    remain very important open problems. Using Hartree-Fock or trial wavefunctions
    approximation, many authors \cite{wigner} proposed different kinds of translational symmetry
    breaking ground states as candidates of the intermediate state .
    In a recent preprint \cite{physics}, I constructed an effective action  to study the
    instability driven by magneto-roton minimum collapsing at a finite wavevector in the pseudo-spin
    channel. I showed that the instability leads to a pseudo-spin density wave (PSDW) with a square lattice
    structure at some intermediate distances. The theory puts the ESF state and
    the PSDW state on the same footing and describe the universality
    class of the quantum phase transition between the two states.
    In the presence of disorders, the properties of the PSDW are
    consistent with all the experimental observations \cite{drag} in the intermediate distances.

   %  They and other authors also studied the bi-layer system from composite boson theory \cite{jap,rev}.
%  Starting from this EPQFM picture, several groups investigated $ I-V $ curves
%  in the presence of small tunneling\cite{balents}.

     In this paper, we use the CB theory developed in \cite{physics} to study the effects of imbalance in BLQH.
     This theory puts spin and charge degree freedoms in the same footing
     and can also be used to explicitly bring out
     the spin-charge connection and classify all the possible
     excitations in a systematic way. As shown in
     \cite{physics}, there are two critical distances and three phases in balanced
     case: $ 0< d < d_{c1} $, the  system is in the ESF, $ d_{c1} <
     d < d_{c2} $, it is in the PSDW with square lattice structure, $
     d > d_{c2} $, it melts into two weakly coupled FL.
     We find that in the ESF side, the imbalance is irrelevant, but the merons carry
     continuously changing fractional charges. While
     in the PSDW side, we map the square lattice PSDW at the balanced
     case into a hard core bosons hopping on a square lattice at half
     filling with long range interactions, then adding a small imbalance in the BLQH corresponds to
     adding a small chemical potential in the boson model. Through this mapping, we find
     the imbalance drives two quantum phase transitions:
     the first one is from the commensurate PSDW state to
     an incommensurate PSDW (IC-PSDW), the second one is from IC-PSDW state to
     the ESF state. Both transitions are first order transitions.
     We compare our phase diagram with the recent experimental data \cite{imbexp}.
     We also discuss briefly the effects of disorders and compare our results
     with previous results achieved from microscopic calculations.

   Consider a bi-layer system with $ N_{1} $ ( $ N_{2}  $ ) electrons in top ( bottom ) layer and
  with interlayer distance $ d $ in the presence of magnetic field $ \vec{B} = \nabla \times \vec{A} $:
\begin{eqnarray}
   H & = & H_{0} + H_{int}    \nonumber  \\
   H_{0} & = &  \int d^{2} x c^{\dagger}_{\alpha}(\vec{x})
   \frac{ (-i \hbar \vec{\nabla} + \frac{e}{c} \vec{A}(\vec{x}) )^{2} }
      {2 m }  c_{\alpha}(\vec{x})                \nonumber   \\
    H_{int} & = &  \frac{1}{2} \int d^{2} x d^{2} x^{\prime} \delta \rho_{\alpha} (\vec{x} )
               V_{\alpha \beta} (\vec{x}-\vec{x}^{\prime} )  \delta \rho_{\beta } ( \vec{x^{\prime}} )
\label{first}
\end{eqnarray}
  where electrons have {\em bare} mass $ m $ and carry charge $ - e $, $ c_{\alpha}, \alpha=1,2 $ are
  electron  operators in top and bottom layers, $ \delta \rho_{\alpha}(\vec{x}) = c^{\dagger}_{\alpha} (\vec{x})
  c_{\alpha} (\vec{x} ) - \bar{\rho}_{\alpha}, \alpha=1,2 $ are normal ordered electron densities on each layer.
  The intralayer interactions
  are $ V_{11}=V_{22}= e^{2}/\epsilon r $, while interlayer interaction is $ V_{12}=V_{21}= e^{2}/ \epsilon
  \sqrt{ r^{2}+ d^{2} } $ where $ \epsilon $ is the dielectric constant.

     Performing a singular gauge transformation $
  \phi_{a}(\vec{x}) = e ^{ i \int d^{2} x^{\prime} arg(\vec{x}-\vec{x}^{\prime} )
  \rho ( \vec{x}^{\prime} ) } c_{a}( \vec{x}) $
     where $ \rho ( \vec{x} ) = c^{\dagger}_{1}( \vec{x} ) c_{1}( \vec{x} ) +
     c^{\dagger}_{2}( \vec{x} ) c_{2}( \vec{x} )  $ is the total density of the bi-layer system.
     We can transform the Hamiltonian Eqn.\ref{first} into a Lagrangian of the Composite
     Boson $ \phi_{a} $ coupled to a Chern-Simon gauge field $ a_{\mu} $ \cite{cbtwo}.
     We can write the two bosons in terms of magnitude and phase $
  \phi_{a}= \sqrt{ \bar{\rho}_{a} + \delta \rho_{a} } e^{i \theta_{a}
  } $, then after absorbing the external gauge potential $ \vec{A} $ into $
   \vec{a} $, we get the Lagrangian in the Coulomb gauge:
\begin{eqnarray}
  {\cal L}  &  =  & i \delta \rho_{+} ( \frac{1}{2} \partial_{\tau} \theta_{+}-  a_{0} ) +
          \frac{ \bar{\rho} }{2m} [ \frac{1}{2} \nabla \theta_{+} + \frac{1}{2} (\nu_{1}-\nu_{2} ) \nabla \theta_{-}
          - \vec{a} ]^{2}
                      \nonumber  \\
    & +  & \frac{1}{2} \delta \rho_{+} V_{+} (\vec{q} )  \delta \rho_{+}
          - \frac{ i }{ 2 \pi} a_{0} ( \nabla \times \vec{a} )
                       \nonumber   \\
    & + & \frac{i}{2} \delta \rho_{-}  \partial_{\tau} \theta_{-} +
          \frac{ \bar{\rho} f }{2m}  ( \frac{1}{2} \nabla \theta_{-} )^{2}
          + \frac{1}{2} \delta \rho_{-} V_{-} (\vec{q} )  \delta \rho_{-}
          - h_{z} \delta \rho_{-}
\label{main}
\end{eqnarray}
     where $ \delta \rho_{+} = \delta \rho_{1} + \delta \rho_{2},
     \delta \rho_{-} = \rho_{1} - \rho_{2}, \theta_{\pm} = \theta_{1} \pm \theta_{2} $,
     they satisfy commutation relations
     $ [ \delta \rho_{\alpha} ( \vec{x} ), \theta_{\beta}( \vec{x}^{\prime} ) ]
         = 2 i \hbar \delta_{\alpha \beta} \delta( \vec{x}-\vec{x}^{\prime} ),
     \alpha, \beta=\pm $.
     $ \bar{\rho} = \bar{\rho_{1}} + \bar{\rho_{2}} $
     and  $ f= 4 \nu_{1} \nu_{2} $ which is equal to 1 at the balanced case,
     $ V_{\pm}= \frac{ V_{11} \pm V_{12} }{2} $,
     $ h_{z}= V_{-} \bar{\rho}_{-} = V_{-} ( \bar{\rho}_{1}-  \bar{\rho}_{2} ) $ plays a role of a Zeeman field.

{\sl (1) Off-diagonal algebraic order and Spin-wave excitation:}
Neglecting vortex excitations, dropping
   a linear derivative term in $ \theta_{-} $ which is irrelevant in the
   ESF state and expanding
   the second term in Eqn.\ref{main} which includes the coupling between the spin sector
   and the charge sector, we can get the complete forms of the three propagators \cite{cbtwo}:
   $ < \theta_{+} \theta_{+} >, < \theta_{-} \theta_{-} >, < \theta_{+} \theta_{-} > $.

    Performing the frequency integral of the $ + $ propagator carefully,
    we find the  {\em leading term} of the equal time correlator of $ \theta_{+} $ stays the same
    as the balanced case $
     < \theta_{+}( - \vec{q} ) \theta_{+} ( \vec{q} ) >  =
     2 \times \frac{2 \pi}{ q^{2} }  + O( \frac{1}{ q } ) $
    which leads to the same off-diagonal algebraic order exponent $ 2 $ as in the balanced case.
%    This fact that the algebraic order in the charge sector is independent of the imbalance
%    maybe expected, because the total filling factor $ \nu_{T}=1 $ stays the same.
   In the $ q, \omega \rightarrow 0 $ limit, we can extract the leading terms in
   the $ \theta_{-} \theta_{-} $ propagator $
      < \theta_{-}  \theta_{-} >  =
      \frac{ 4 V_{-}(q) }{ \omega^{2} + f  v^{2} q^{2} } $
    where $ v $ is the spin wave velocity in the balanced case and
    we can identify the spin wave velocity in the im-balanced case $
     v^{2}_{im} = f v^{2} = 4 \nu_{1} \nu_{2} v^{2} = 4 \nu_{1}( 1-\nu_{1} )
     v^{2} $ which shows that the spin-wave velocity attains its maximum at the
     balanced case and decreases parabolically as im-balance increases.

%  Similarly, in $ q, \omega \rightarrow 0 $ limit, we can also extract the leading terms in
%  the $ \theta_{+} \theta_{-} $ propergator $ < \theta_{+}  \theta_{-} >
%      = -( \nu_{1}-\nu_{2} ) < \theta_{-}  \theta_{-} > $
%   which shows that the behavior of $ < \theta_{+}  \theta_{-} > $ is dictated by that of
%    $  < \theta_{-}  \theta_{-} >  $ instead of  $  < \theta_{+}  \theta_{+} >  $.
%   When the vortex excitations to be discussed in the following are included,
%   the spin wave velocity will be renormalized down,
    As analyzed in detail in \cite{cbtwo},
    it is hard to incorporate the Lowest Landau Level (LLL) projection in the CB approach,
    so the spin wave velocity
    can only be taken as a phenomenological parameter to be fitted to
    experiments. But as shown explicitly in the next subsection, the CB theory can be used
    to classify topological excitations correctly and systematically .

{\sl (2) Topological excitations:} There are following 4 kinds of
topological excitations:
   $ \Delta \theta_{1} = \pm 2 \pi,
  \Delta \theta_{2} = 0 $ or $ \Delta \theta_{1} =0, \Delta \theta_{2} = \pm 2 \pi $.
  Namely $ ( m_{1}, m_{2} ) = ( \pm 1, 0 ) $ or  $ ( m_{1}, m_{2} ) = ( 0, \pm 1 ) $ where $ m_{1} $
  and $ m_{2} $ are the winding numbers in $ \theta_{1} $ and $ \theta_{2} $ respectively.
  The fractional charges can be determined from the constraint
  $ \nabla \times \vec{a} = 2 \pi \delta \rho $ and the finiteness
  of the energy in the charge sector:
\begin{eqnarray}
  q &= & \frac{1}{ 2 \pi} \oint  \vec{a} \cdot d \vec{l}
  = \frac{1}{ 2 \pi} \times \frac{1}{2} \oint [ \nabla \theta_{+} + ( \nu_{1} -\nu_{2} ) \nabla \theta_{-} ]
   \cdot d \vec{l}
    \nonumber  \\
  & = &  \frac{1}{2} [ m_{+} + ( \nu_{1}- \nu_{2} ) m_{-} ]
\end{eqnarray}
   where $ m_{\pm}= m_{1} \pm m_{2} $.

  We can classify all the possible topological excitations in terms of $ ( q, m_{-} ) $ in the
  following table:

\vspace{0.25cm}

%\begin{table}
\begin{tabular}{ |c|c|c|c|c| }
   $ (m_{1}, m_{2} ) $      &  $ (1,0 ) $   & $ (-1,0) $  & $ (0,1) $  &   $ ( 0, -1)  $      \\  \hline
  $ m_{-} $  &  $ 1 $        &   $ -1 $    &  $ -1 $        &  $  1 $     \\   \hline
  $ m_{+} $  &  $ 1 $        &   $ -1 $    &  $ 1 $        &  $  -1 $     \\   \hline
   $ q    $  &  $ \nu_{1} $        &   $ -\nu_{1} $    &  $ \nu_{2} $        &  $  -\nu_{2} $
\end{tabular}
\par
\vspace{0.25cm} {\footnotesize  {Table 1: The fractional charge in
im-balanced case } }
%\end{table}
\vspace{0.25cm}
%\caption{ The factional charges in im-balanced case }
%\end{table}

% We can see that any deconfined excitations
% with  $ ( m_{-} =0 , m_{+} = 2m ) $  have charges $ q=  \frac{1}{2} m_{+}=m $ which
% must be integers. While the charge of  $ ( m_{+}=0, m_{-}=2m  ) $ is $ q= ( \nu_{1}- \nu_{2} ) m $
% which is charge neutral only at the balanced case.

  The finiteness of the energy in the spin sector dictates that
  any finite energy excitation must have $ m_{-}=0 $, so
  the merons listed in table 1 are confined into the following two possible pairs at low temperature.
   (1)  Charge neutral pairs: $ (\pm \nu_{1}, \pm 1 ) $ or $ (\pm \nu_{2}, \mp 1 ) $.
        The NGM will turn into these charge neutral pairs at large wavevectors.
        The pairs behave as bosons.
   (2)  Charge $ 1 $ pair $ (\nu_{1}, 1 ) + (\nu_{2}, -1 ) $ or charge $ -1 $ pair
        $ (- \nu_{1}, -1 ) + (- \nu_{2}, 1 ) $. The pairs behave as fermions.
        They maybe the lowest charged excitations in BLQH and the main dissipation sources
        for the charge transports.

{\sl  (3) Im-balance driven quantum phase transitions in the PSDW
          side:}
    As said in (1), the imbalance is irrelevant in the ESF
    side, but we expect it is important in the PSDW side. Starting
    from the ESF side, it would be useful to calculate how the mageto-roton minimum
    depends on the imbalance \cite{jog,cbtwo}, unfortunately this is beyond the scope
    of the CB theory. Here we take a different strategy: starting
    from the PSDW side and studying how a small imbalance affects the PSDW.
    If the imbalance is sufficiently small,
    we expect the C-PSDW at the balanced case in Fig.2a is a very good reference state.
    Because it is a square lattice with the "up" pseudo-spins
    taking sublattice $ A $ and the "down" pseudo-spins taking
    sublattice $ B $, it is reasonable to start from a lattice model from the PSDW
    side. If we think the PSDW as a charge density wave ( CDW )  of bosons at half filling
    on a square lattice, then we can view the ESF to the PSDW as a
    superfluid to CDW transition in a boson Hubbard model of {\em hard core} bosons  {\em near } half filling
    hopping on square lattice of bosons:
\begin{eqnarray}
   H  & = & -t \sum_{ < ij > } ( b^{\dagger}_{i} b_{j} + h.c. )
          - \mu \sum_{i} n_{i}    \nonumber   \\
    & + &  V_{1} \sum_{ <ij> } ( n_{i}-1/2 )( n_{j}-1/2)
      +   V_{2} \sum_{ <<ik>> } n_{i} n_{k} + \cdots
\label{boson}
\end{eqnarray}
    where  $ S^{+} = b^{\dagger}=c^{\dagger}_{1} c_{2}, S^{-}= b = c^{\dagger}_{2} c_{1},
     S^{z}= \frac{1}{2}( c^{\dagger}_{1} c_{1} - c^{\dagger}_{2} c_{2} )=  b^{\dagger} b-1/2 $
     are the pseudo-spin density and boson operators.
    At the total filling factor $ \nu_{T}=1 $, we can impose the local constraint
    $ c^{\dagger}_{1} c_{1} + c^{\dagger}_{2} c_{2} =1 $.
    $ n_{i} = b^{\dagger}_{i} b_{i} $ is the boson density, $ t $ is the
    nearest neighbor hopping
    amplitude, $ V_{1}, V_{2} $ are the nearest and next nearest neighbor repulsive
    interactions between the bosons. The $ \cdots $ may include further neighbor interactions.
    Because of the long-rang Coulomb interaction in Eqn.\ref{main},
    it is important to keep all the long-range interactions in
    the lattice model Eqn.\ref{boson}.
    If the chemical potential $ \mu=0 $, the bosons are at the half filling $ < n_{i} >=1/2
    $ which corresponds to the balanced case $ \nu=1/2 $. The
    particle-hole symmetry of Eqn.\ref{boson} corresponds to the $ Z_{2} $ exchange
    symmetry of the BLQH.
    If the chemical potential $ \mu \neq 0 $, the bosons are slightly away from the half filling
    which corresponds to the slightly imbalanced case.

    The boson Hubbard model Eqn. \ref{boson} in square lattice
    at generic commensurate filling factors $ f=p/q $ ( $ p, q $ are relative prime numbers ) were
    systematically studied in \cite{pq1}
    by performing the charge-vortex duality transformation.
    Recently, we applied the dual approach to study reentrant
    supersolids and quantum phase transitions from solids to the
    reentrant supersolids on extended boson Hubbard models Eqn.\ref{boson} at in-commensurate filling factors
    in bipartite lattices such as square and  honeycomb lattices.
    In the following,
    we apply the results achieved in \cite{nature} to the present
    problem on square lattice at and slightly away from half filling. At $ q=2 $, there
    are two dual vortex fields $ \psi_{a} $ and $ \psi_{b} $.
   Moving {\em slightly} away from half filling $ f=1/2 $ corresponds to adding
   a small {\em mean} dual magnetic field $ H \sim  \delta f= f-1/2 $ in the dual action.
   The most general action invariant under all the MSG transformation laws upto quartic
   terms is \cite{pq1,nature}:
\begin{eqnarray}
    {\cal L} & = & \sum_{\alpha=a/b} | (  \partial_{\mu} - i A_{\mu} ) \psi_{\alpha} |^{2} + r | \psi_{\alpha} |^{2}
    +  \frac{1}{4} ( \epsilon_{\mu \nu \lambda} \partial_{\nu} A_{\lambda}
    - 2 \pi \delta f \delta_{\mu \tau})^{2}   \nonumber  \\
       & +  & \gamma_{0} ( | \psi_{a} |^{2} + |\psi_{b} |^{2} )^{2} -
                       \gamma_{1} ( | \psi_{a} |^{2} - |\psi_{b} |^{2} )^{2} + \cdots
\label{away}
\end{eqnarray}
     where $ A_{\mu} $ is a non-compact  $ U(1) $ gauge field. Upto
     the quartic level, Eqn.\ref{away} is the same in
     square lattice and in honeycomb lattice.
     If $ r > 0 $, the system is in the superfluid state $  < \psi_{l} > =0 $ for every $ l=a/b $.
     If $ r < 0 $, the system is in the insulating state $ < \psi_{l} > \neq 0 $ for
     at least one $ l $.  We assume $ r < 0 $ in Eqn.\ref{away}, so the system is in the insulating state.
     $ \gamma_{1} > 0 $ ( $ \gamma_{1} < 0 $ ) corresponds to the
     Ising ( or Easy-plane ) limit. The insulating state takes the CDW  state ( or valence
     bond solid (VBS) state ).

     In the balanced case $ \delta f=0 $, the SF to the VBS transition
     in the easy plane limit was argued to be 2nd order through a novel deconfined quantum
     critical point \cite{pq1}. However, the boson Hubbard model
     Eqn.\ref{boson} on the PSDW side
     corresponds to the Ising limit in the dual model Eqn.\ref{away}, therefore $ \gamma_{1} > 0 $.
     The SF to the CDW transition in the Ising limit
     is first order. This is consistent
     with the first order ESF to PSDW transition driven by the collapsing of magnetoroton minimum studied in
     \cite{physics}.
%    As shown in \cite{physics}, although the
%     system is in the easy-plane limit in the $ q \rightarrow 0 $
%     limit, but it turns into the Ising limit at the lattice scale $
%     q \sim q_{0} \sim 1/a $. In the ENS, because we are working on
%     the lattice scale, so the boson Hubbard model Eqn.\ref{boson}
%     is at the Ising limit, therefore $ \gamma_{1} > 0 $.
%     So the SF to the CDW is still first order,

    In the CDW order side, the mean field solution is $ \psi_{a}=1, \psi_{b}=0 $ or vice versa.
    In the slightly imbalance case $ \delta f \neq 0 $,
    setting $ \psi_{b}=0 $ in Eqn.\ref{away} leads to:
\begin{eqnarray}
  {\cal L} & = & | (  \partial_{\mu} - i A_{\mu} ) \psi_{a} |^{2} + r | \psi_{a}
  |^{2} + \gamma_{0} | \psi_{a} |^{4} + \cdots     \nonumber  \\
     & + & \frac{1}{4} ( \epsilon_{\mu \nu \lambda} \partial_{\nu} A_{\lambda}
    - 2 \pi \delta f \delta_{\mu \tau})^{2}
\label{is}
\end{eqnarray}
   where $ \rho_{A} =  \psi^{\dagger}_{a} \psi_{a} $ should be interpreted as the
   vacancy number, while the vortices in its phase winding  are interpreted as  boson number.
   Of course, a negative imbalance can simply achieved by a particle
   hole transformation $ \psi_{a} \rightarrow \psi^{\dagger}_{a},
   \delta f \rightarrow - \delta f $ in Eqn.\ref{is}.

   Eqn.\ref{is} has the structure identical to the conventional $ q=1 $ component
   Ginzburg-Landau model for a  type II " superconductor "  in a "magnetic"
   field. It was well known that as the magnetic field increases, there are
   two {\em first order} phase transitions: $ H < H_{c1} $, the system is in the
   Messiner phase, $ H_{c1} < H < H_{c2} $, it is in the vortex
   lattice phase, $ H > H_{c2} $ it is in the normal phase.
   In the present boson problem with the nearest neighbor interaction $
   V_{1} > 0 $ and further neighbor interactions in Eqn.\ref{boson} which stabilizes
   the CDW state at $ f=1/2$ ( Fig.2a ) and IC-CDW state at $ f=1/2 + \delta f $ ( Fig.2b),
   this corresponds to C-CDW to IC-CDW to
   superfluid transition shown in Fig.1a. Transferring back to the
   original BLQH problem, the small imbalance will first drive the C-PSDW to
   the In-commensurate pseudo-spin density wave (IC-PSDW), then drive a 1st order transition from
   the IC-PSDW to the ESF shown in Fig.1b.

\vspace{0.25cm}

\epsfig{file=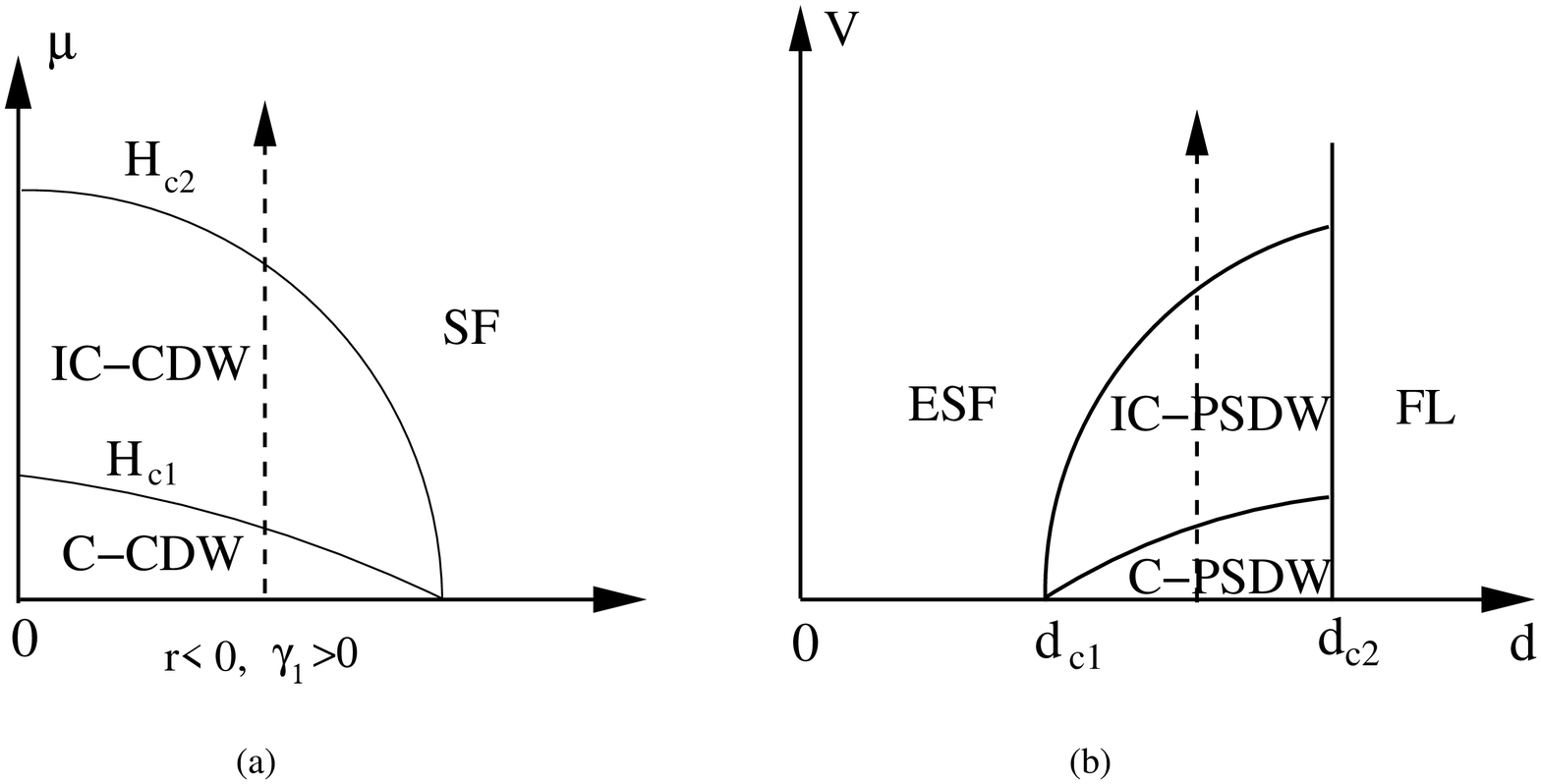,width=3.2in,height=1.6in,angle=0}

\vspace{0.25cm}

{\footnotesize {\bf Fig.1:} (a) The phase diagram of the boson
      Hubbard model Eqn.\ref{boson} slightly away from the half filling.
      $ \mu $ is the chemical potential. (b) The bias voltage $  V $ versus
      distance $ d $ phase diagram at zero temperature. IC-PSDW
      stands for the incommensurate PSDW. The dashed line
      is the experimental path investigated in \cite{imbexp}. All the transitions are first order
      transitions. }

\vspace{0.25cm}

    As shown in Fig.1b, the bias voltage increases, the imbalance will first
    introduce interstitials in the top layer and vacancies in the
    bottom layer, namely, turn the C-PSDW into the IC-PSDW whose charge distributions are shown in
    Fig.2a and 2b respectively, then the whole IC-PSDW melts into the ESF through a 1st
    order transition.

\vspace{0.25cm}

\epsfig{file=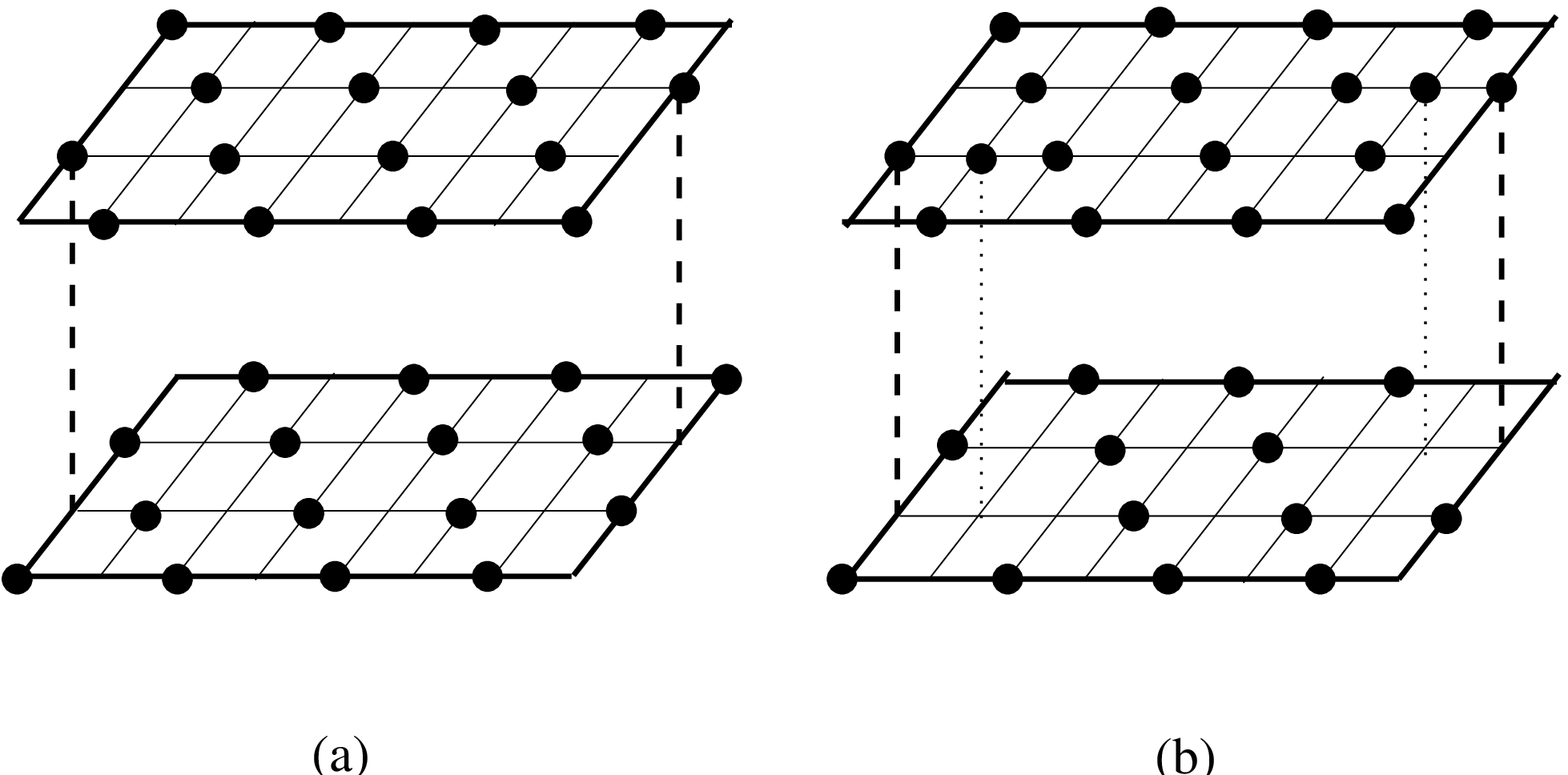,width=3.0in,height=0.8in,angle=0}

\vspace{0.25cm}

{\footnotesize {\bf Fig.2:} (a) The charge distribution of the PSDW
in a square lattice. The dashed line stand for "up" pseudo-spins
which take sublattice A, while "down" pseudo-spins take sublattice
B. (b) The  charge distribution of the IC-PSDW in a square lattice.
The dotted line stand for "inverted" pseudo-spins which play the
role of interstitials in the top layer or vacancies in the bottom
layer. The number of interstitials in the top layer is equal to that
of vacancies in the bottom layer.}

\vspace{0.25cm}

    Disorders may smear all the 1st order
  transitions in Fig.1b into 2nd order transitions.
  The disorders may also transfer the long
  range lattice orders of the C-PSDW and IC-PSDW into short range ones. The fact make the
  observation of the commensurate and incommensurate lattice structures by light scattering experiment \cite{light}
  difficult.

   The dashed line in Fig.1b was investigated in a recent experiment
   \cite{imbexp}. But the first phase transition in Fig.1b was not paid attention in the
   experiment where the phase diagram was drawn against fixed charge imbalance $
   h_{z} $ instead of fixed bias voltage $ V $.
   So the C-PSDW phase was crushed into the horizontal axis.
   A simple mean field argument leads to the linear scaling of the second
   transition line $ V \sim d-d_{c1} $.
   A parabolic behavior $ h^{2}_{z} \sim d-d_{c1} $ was found for the
   shape of the second transition at very small imbalances. We
   expect the disorders may transform the linear behavior to the
   parabolic one. In the presence of disorders, all the properties of the C-PSDW and IC-PSDW
   are consistent with the experimental
   observations in \cite{imbexp} on the intermediate phase at small imbalances.
%   Unfortunately,
%   the effective CB theory can not be used to explain the parabolic
%   dependence.

  When the distance of the two layers is further increased to larger than a second critical
  distance $ d_{c2} $, then all the signature of the interlayer
  coherent state are lost, the two layers are decoupled into two separate $ \nu=1/2 $ CF Fermi liquid state ( Fig.1b ).
  We expect that there is a level crossing and associated first order transition at $ d_{c2} $.
  When $ d > d_{c2 } $, increasing the bias voltage may not transform the two decoupled
  FL state back into the ESF.

{\sl (4) Discussions:}
    By projecting into the LLL and then performing the HF approximation
    ( LLL+HF approach ), the authors in \cite{jog} found that the transition at
    $ h_{z}=0, d=d_{c1} $ is an instability through a 1st order transition
    to a pseudospin density wave state driven by the gap closing of magneto-roton minimum
    at a finite wave-vector. Starting from the ESF side, their numerical results indicated that
    the imbalance increases the spin stiffness and also the critical
    distance $ d_{c1} $.
%   In this LLL+HF approach, the charge sector was integrated out, therefore may not
%   address the interplay between the QH effects in the charge sector and the interlayer phase
%   coherence in the spin sector.
%   The charge sector also becomes gapless at $ d_{c1} $,
%   so its fluctuation can not be ignored near $ d_{c1} $.
    It is not known if the LLL+HF calculations are accurate in
    describing the transition from the ESF to the pseudo-spin
    density wave state.
%    studying quantum Hall
%    states in SLQH
%    or BLQH systems with a bulk gap, but its accuracy may drop considerably
%   in the BLQH with a broken symmetry state and its associated gapless Goldstone
%    mode.
%    All the calculations in LLL+HF approach assumes
%    that the ground state wavefunction at any finite $ d $ is still the  (111) wavefunction.
%    However as shown in \cite{wave,square}, the wavefunction at any finite $ d $ is {\em qualitatively}
%    different from  the (111) wavefunction which is good only at $ d=0 $. So the excitation spectra
%   in the ESF side calculated by the LLL+HF based on the (111) wavefunction may not have quantitatively correct
%    distance dependence.
    Our effective theory circumvents the difficulty
    associated with the still not precisely known wavefunction at any finite $ d $ \cite{wave,square}. We
    study the effects of small imbalance from the PSDW side and
    map its effect as a chemical potential of hard core bosons
    with long range interactions
    hopping on a square lattice near half filling, namely, mapping Fig.1a to Fig.1b. Our upper phase boundary
    in Fig.1b is consistent with that achieved in \cite{jog} from the ESF side by the LLL+HF calculation. We also worked
    out the lattice structures of the C-PSDW and IC-PSDW
    and the whole physical picture  along the dashed line in
    Fig.1b. It is not known how to apply the LLL+HF theory in
    \cite{jog} to study
    the PSDW side. The effective theory presented in this paper is complementary to and goes well beyond the
    previous LLL+HF calculations.
%    As stated previously,
%    it is hard to incorporate the LLL projection in the CB approach.
%    So the effective approach in this paper and the LLL+HF approach in \cite{jog} are complementary to each other.

%   In the presence of disorder, the DL state may also become a glassy phase like Dipolar Glass ( DG )
%   state where only short range of crystalline order should survive.
%   The disorder may also smear the first order
%   transition at $ d_{c2} $ into a 2nd order transition.

% The shape of the phase boundary
% matches qualitatively that calculated by HF approximation.

    I thank J. K. Jain, P. A. Lee and A. H. Macdonald for helpful discussions.
    I also thank Yan Chen, Z. D. Wang and F.C. Zhang for hospitality during my visit in Hong
    Kong University in writing this paper.

\end{multicols}
\end{document}